\documentclass[aps, reprint,onecolumn,amsmath,amssymbsuperscriptaddress,preprintnumbers,nofootinbib]{revtex4-1}
\pdfoutput=1
\usepackage{graphicx}
\usepackage{amssymb}
\usepackage{amsmath}
\usepackage{amsfonts}
\usepackage{latexsym}
\usepackage{url}
\usepackage[all]{xy}
\usepackage{slashed}
\usepackage{hyperref}
\newcommand{\be}{\begin{eqnarray}}
\newcommand{\ee}{\end{eqnarray}}

\newcommand{\keV}{\ensuremath{\mathrm{keV}}}

\newcommand{\GeV}{\ensuremath{\mathrm{GeV}}}

\newcommand{\kg}{\ensuremath{\mathrm{kg}}}

\newcommand{\cpd}{\ensuremath {\mathrm{cpd}}}

\newcommand{\Kforty}{\ensuremath{{}^{40}\mathrm{K}}}

\newcommand{\Arforty}{\ensuremath{{}^{40}\mathrm{Ar}}}

\begin{document}

\title{A reply to the criticism of our work
  (\href{http://arxiv.org/abs/arXiv:1210.5501}{arXiv:1210.5501}) by
  the DAMA collaboration} \title{Addendum to\\``On an unverified nuclear decay and its role in the DAMA
  experiment''}
\author{Josef Pradler}
\email{jpradler@pha.jhu.edu} \affiliation{Department of Physics and
  Astronomy, Johns Hopkins University, Baltimore, MD 21218, USA}
\author{Itay Yavin} \email{iyavin@perimeterinstitute.ca}
\affiliation{Perimeter Institute for Theoretical Physics 31 Caroline
  St. N, Waterloo, Ontario, Canada N2L 2Y5} \affiliation{Department of
  Physics \& Astronomy, McMaster University 1280 Main St. W, Hamilton,
  Ontario, Canada, L8S 4L8}

\begin{abstract}
  We reply to the critiques of our paper
  \href{http://arxiv.org/abs/arXiv:1210.5501}{arXiv:1210.5501} by the
  DAMA collaboration which appeared in
  \href{http://arxiv.org/abs/arXiv:1210.6199}{arXiv:1210.6199} and
  \href{http://arxiv.org/abs/arXiv:1211.6346}{arXiv:1211.6346}. Our
  original claim that the observed background levels are likely to
  require a large modulation fraction of any putative signal holds. In
  fact, in light of DAMA's recent comment our claim is further
  corroborated. We identify the source of the discrepancy between our
  own analysis and DAMA's claimed levels of unmodulated
  background. Our analysis indicates that the background in the signal
  region as reported by DAMA is indeed likely underestimated.
\end{abstract}

\maketitle

In a recent publication~\cite{Pradler:2012qt} we pointed out that the
empirical verification of a certain special decay branch of \Kforty\
remains outstanding and discussed the importance of such a
measurement. In addition, we discussed the general role \Kforty\ plays
as an important background in the DAMA
experiment~\cite{Bernabei:2010mq}. Shortly after our paper appeared,
the DAMA collaboration has criticized some of its
findings~\cite{Bernabei:2012ab}. We replied in a previous version~(v1)
of this manuscript which prompted the collaboration to post another
comment in~\cite{Bernabei:2012rc}. To help everybody else to keep
better track of the discussion, we address both comments by the DAMA
collaboration,~\cite{Bernabei:2012ab} and~\cite{Bernabei:2012rc}
jointly. All our previous statements remain intact.
There are three points of substance in~\cite{Bernabei:2012ab}
and~\cite{Bernabei:2012rc}:

\begin{description}
\item[{\bf Critique 1}] In~\cite{Bernabei:2012ab} the collaboration
  states that the contribution of this special decay of \Kforty\ to
  the total \Kforty\ contribution is only about 10\%. They therefore
  criticize our discussion of this branch as "captious". In addition,
  in~\cite{Bernabei:2012rc} it is claimed that our calculation for the
  theoretical prediction of the \Kforty\ decay directly to the ground
  state of \Arforty\ is not performed correctly because the branching
  from the K-shell had not been taken into account.
\item[{\bf Reply}] Our calculation of the ratio of K-shell electron capture rate to
  the $\beta^+$ emission rate is correct and does not miss the K-shell branching because
  we make direct use of the K-shell atomic wave-function. The mention
  of higher-shell captures by the collaboration is captious: these
  transitions are subdominant relative to the K-shell capture, $\sim
  10\%$ (much smaller than the overall uncertainty on the ratio); they
  also contribute to energy depositions only below threshold.
  More importantly, the total decay of \Kforty\ is a serious
  background in the DAMA experiment and it was in fact one of our
  findings that this special branch is subdominant to the one where
  \Kforty\ decays into the \Arforty\ excited state, followed by a
  1.46~MeV gamma-ray which escapes detection. In their
  comment~\cite{Bernabei:2012ab} DAMA quotes (for the first time in
  print) that the former contribution is 10\% of the total $\Kforty$
  contribution.
  This is indeed in good agreement with what we found using the
  theoretical branching ratio and using the independent Monte Carlo
  simulation of Ref.~\cite{Kudryavtsev:2010zza}. This agreement gives further
  credence to our treatment of potassium decays in DAMA.

\item[{\bf Critique 2}] In DAMA's reply~\cite{Bernabei:2012ab} to our
  paper the collaboration states that the content of
  $^{\mathrm{nat}}$K has been measured by investigation of the double
  coincidence, that it is independent of the branching ratio into the
  ground state, and that its average value over all crystals is
  13~ppb. Statements based on 20~ppb are therefore
  wrong. In~\cite{Bernabei:2012rc} the collaboration reminds us that
  the number~13~ppb is published in~\cite{Bernabei:2009pg} and refutes
  our critique that no discussion supporting this number is available
  would be unfounded and not justified.
\item[{\bf Reply}] No discussion or mentioning of the average \Kforty\
  contamination of 13~ppb can be found in the published TAUP
  conference proceedings~\cite{Bernabei:2009pg}. Neither could we
  locate this number in any other of DAMA's published
  works. The average contamination of 13~ppb can indeed
  be found on slide number~8 in the TAUP talk by
  Nozzoli~\cite{Nozzoli}. Importantly, however, no data is presented
  to support this number (such as \textit{e.g.}~the individual
  crystal-by-crystal contaminations,) nor does the collaboration
  provide the uncertainty associated with it.
  May that as it be, our results are presented for a whole range of
  $^{\mathrm{nat}}$K contaminations between \mbox{1-100}~ppb including
  the DAMA quoted maximum individual crystal contamination of 20~ppb,
  which appeared in the official publication describing the DAMA
  apparatus~\cite{Bernabei:2008yh}. Using 13~ppb we find that the
  required signal modulation fraction is above 20\%.
  In addition, in~\cite{Pradler:2012qt} we identified a way to measure
  the contamination level in a manner that is free from the Monte
  Carlo modeling required by the coincidence method.
  Considering the immense importance of the precise amount of
  potassium background to any interpretation in terms of a dark matter
  signal the lack of details so far provided by the collaboration is
  unsettling.

\item[{\bf Critique 3}] Finally, in their reply~\cite{Bernabei:2012ab}
  to our paper~\cite{Pradler:2012qt} the DAMA collaboration claims an
  upper limit of $S_{0}\leq 0.25$~cpd/kg/keV for the unmodulated part of the
  signal in the 2-4~keV energy bin.
  Given a residual rate of
  $0.0194\pm0.0022$~cpd/kg/keV~\cite{Bernabei:2010mq}, this would
  allow for a modulation fraction of 6-10\% which can be accommodated
  with many models of Dark Matter~\cite{Bernabei:2008yi}, contrary to
  our conclusions.
  Our conclusions are based on the assumption of a flat background
  component at the conservative level of $0.85$~cpd/kg/day.
  In the follow-up~\cite{Bernabei:2012rc} the DAMA collaboration
  criticizes this assumption as ``completely arbitrary'' because 1) it
  is not based on the knowledge of background contributions and 2) an
  assessment of backgrounds without accounting for a signal
  contribution $S_0$ in a fit is ``always methodologically
  incorrect.''
  
\item[{\bf Reply}] 
  What regards DAMA's critique of our ``ad-hoc'' assumption of a flat
  background: As argued in our original work~\cite{Pradler:2012qt}, a
  flat background component is expected because it is a universal
  feature of $\beta^-$ decays for small electron velocities and it is
  also typical of low-energy Compton background. The question really
  is: how large is it? Admittedly, this may be very difficult to
  quantify precisely even with the full insight into the
  crystal-by-crystal spectrum as it presents itself to the DAMA
  collaboration.
  However, as we argued in our work, inspection of the
  ``signal-sidebands'' between $5-10$\,\keV\ and $20-40$\,\keV\ (there
  is no published data between $10-20$\,\keV)  strongly supports the notion of a flat
  background contribution at the $0.85$~cpd/kg/day level.
  Given that such flat backgrounds typically span decades in keV
  recoil energy (depending on the Q-value of the parent decay) we do
  not think that a joint fit including a contribution from $S_0$ is
  mandatory for proposing our hypothesis, nor is it going to be
  rejected by it\footnote{However, the reader is referred to the text below and to Fig.~\ref{fig:fullFit} where we address directly the effects of including the signal in the fit.}.
  In fact the value $0.85$~cpd/kg/day in our original work is not the
  result of a fit but is the rate associated with the lowest point (at
  $E=4.6$~keV, see dashed black line in Fig.~\ref{fig:total-fit}).
  Our purpose in \cite{Pradler:2012qt} was to raise awareness that
  such levels of background may very well be present in the
  experiment---and DAMA has not provided a detailed discussion to
  convince us otherwise---hence challenging the DM interpretations
  which come with weak modulation fractions $\lesssim 20\%$.

  What regards the upper limit $S_{0}\leq 0.25$~cpd/kg/keV on the
  unmodulated signal part as mentioned in DAMA's
  comment~\cite{Bernabei:2012ab}: The collaboration says that they
  account for this limit in~\cite{Bernabei:2008yi} when assessing the
  viability of various DM scenarios. Similar to our reply of
  critique~2, we could not find this number or a discussion thereof in
  any of DAMA's published works. Again we have to resort to slide~8 of
  the talk by Nozzoli~\cite{Nozzoli} where the limit is quoted and a
  curve for the background model is reported.  
  In Fig.~\ref{fig:total-fit} we show our reproduction of this
  background, which is a Gaussian centered at 3.2~keV together with a
  linear fit as explained in the Appendix.
  As is clear from the plot, the linearly rising function is not
  supported by the data above 4~keV (the statistical error bars are
  shown but are barely visible.)
  More importantly, the decrease towards lower energies strongly
  supports the notion that the background present in the signal region
  2-4 keV is underestimated.
  In the Appendix we offer details and a further-going discussion of
  this.

\end{description}

What should be clear from this reply is that a more detailed and
careful discussion of backgrounds by the DAMA collaboration is called
for.  That was among the main conclusions in~\cite{Pradler:2012qt}
where we also identified measures which could help to clarify some
of the uncertainties.  We invite the
interested readers to consult the appendix where more details are
presented and to reproduce our results themselves.

\newpage
\clearpage

\begin{figure}[t]
\begin{center}
\includegraphics[width=0.70\columnwidth]{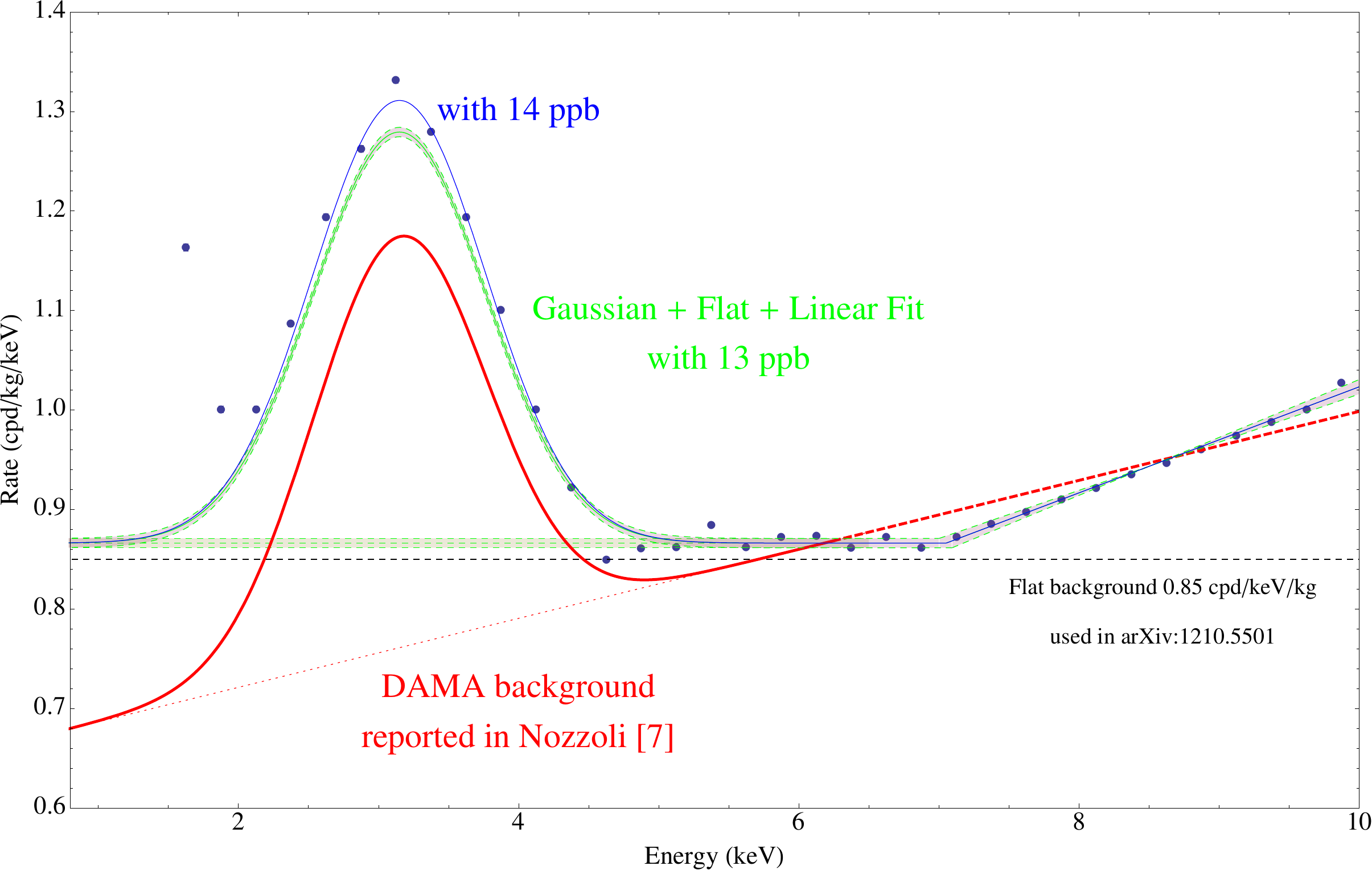}
\end{center}
\caption{The dots (with error-bars) are the single-hit rate reported
  by DAMA in~\cite{Bernabei:2008yh}.  The thick red curve shows our
  reproduction of the background curve presented by DAMA in
  Ref.~\cite{Nozzoli}. The dashed curve from 6.2 keV to 10 keV is the
  continuation of the linear trend. The linear trend is obtained by
  fitting the data between 5.3 keV and 10 keV and results in excellent
  agreement with the reported background~\cite{Nozzoli}, as explained
  in the Appendix. Using the same Gaussian, but with a better model
  for the background above 5~keV (flat + linear), we obtain the solid
  green curve. In blue, we show the resulting curve in the case when
  $^{\mathrm{nat}}$K contamination level is increased to 14 ppb
  instead. We emphasize that we did not fit to the data below 5~keV,
  but used the Gaussian as reported by the collaboration
  in~\cite{Nozzoli}.}
\label{fig:total-fit}
\end{figure}

\appendix


\renewcommand{\theequation}{A-\arabic{equation}}
\setcounter{equation}{0}

\section{Details of the analysis}

The unmodulated background presented on slide 8 of
Nozzoli~\cite{Nozzoli} is reproduced accurately by fitting a linear function
to the data points between 5.3~keV and 10~keV and adding to it a single
Gaussian whose parameters are taken directly from ref.~\cite{Nozzoli},
\be
\label{eqn:Nozzoli_BKG}
\text{BKG}_{\text{Nozzoli}}(E) =
\frac{A}{\sqrt{2\pi\sigma^2}}\exp\left(-\frac{(E-E_{\rm
      K40})^2}{2\sigma^2} \right) + \text{slope}\times E +
\text{intercept} .
\ee 
The amplitude of the Gaussian is $A=0.64$ which is in good agreement
with  the results of~\cite{Kudryavtsev:2010zza} with a
$^{\mathrm{nat}}$K contamination of 13~ppb. The center and spread of the Gaussian are in good agreement with the expected 3.2~keV energy deposition
($E_{\rm K40} = 3.15$~keV) and resolution ($\sigma =0.618$ keV). 
One would naively expect that the quoted upper limit $S_0\lesssim
0.25$~cpd/kg/keV was derived by subtracting this simple background
model from the data. However, in contradiction to what the DAMA collaboration is quoting this yields only $S_0 \approx 0.14$
cpd/kg/keV in the 2-4~\keV\ energy bin as can be easily checked. 

Given the negligible statistical error bars, it is seen in Fig.~\ref{fig:total-fit} that the linear
fit in Eq.~(\ref{eqn:Nozzoli_BKG}) provides an inadequate description
of the data between 5-10~keV where little DM signal is expected.
A flat background up to about 7~keV where it is broken and followed by a linearly rising one provides a much better agreement with the data\footnote{ We have also tried a more general model involving two linear fits, but the results are qualitatively unchanged - the fit between 5-7 keV prefers to be flat.},
\be
\text{BKG}_{\text{linear+flat}}(E) = \left\{ 
\begin{array}{cc}
0.866  & E < 7.05 ~{\rm keV} \\
0.491+0.053 E  & E > 7.05 ~{\rm keV} 
\end{array}
\right.  \ee Such a model is also physically well-motivated. The
background from $\beta^-$-emitters is entirely flat at low
energies. The rise above 7~keV is more difficult to interpret because
the DAMA collaboration has not released the the immediate spectrum
above 10~keV. For example, it could be a broad Iodine escape peak
between 14-18~\keV\ originating from external $^{210}$Pb
contamination. Though such features have been measured previously with
NaI(Tl) crystals~\cite{Cebrian:2002vd}, without further insight into
the spectrum this remains a speculation.

\begin{figure}[bt]
\begin{center}
\includegraphics[width=0.75 \columnwidth]{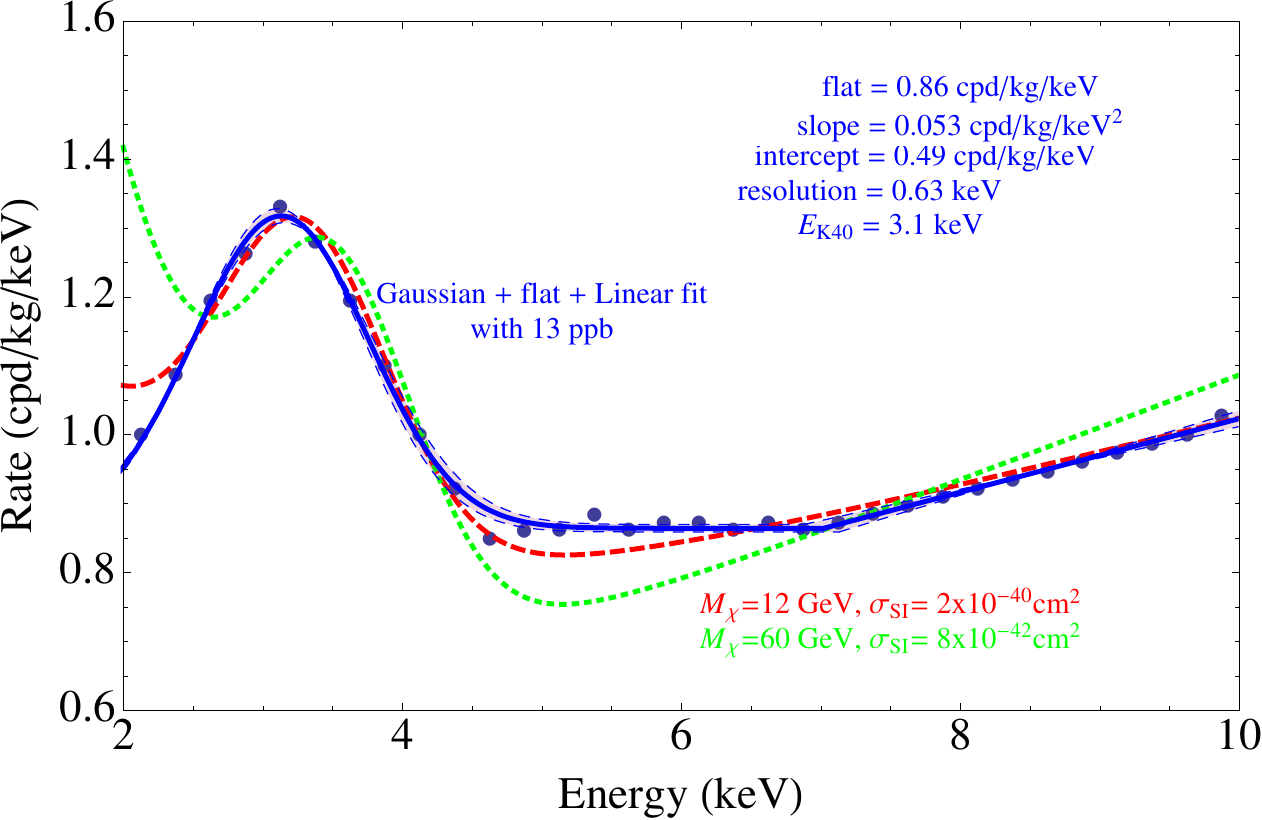}
\end{center}
\caption{The solid-blue line represents a full fit to the unmodulated
  DAMA data composed of a Gaussian component, a flat component, and a
  linearly rising component. The center of the Gaussian, its spread,
  the flat piece, and the slope and intercept of the linear function
  were allowed to float (total of 5 parameters). We stress that the
  amplitude of the Gaussian was fixed to a level which corresponds to
  a 13~ppb of $^{\mathrm{nat}}$K contamination as quoted by the
  collaboration. The dashed-red (dotted-green) is the result of
  fitting the same background model above together with an
  hypothetical dark matter signal corresponding to the best-fit
  low-mass (high-mass) region based on the modulation data with
  parameters $m_{\rm DM} = 12~\GeV$ and $\sigma_{\rm SI} = 2\times
  10^{-40}$ ($m_{\rm DM} = 60~\GeV$ and $\sigma_{\rm SI} = 8\times
  10^{-42}$).}
\label{fig:fullFit}
\end{figure}

We have also verified that a more general fit to the entire data
(including the width and center of the Gaussian) does not result in
any substantial change to the flat component which becomes
0.86~\cpd/\kg/\keV, in agreement with what we used in our original
work~\cite{Pradler:2012qt}. In fact, this model results in a
disturbingly good fit to the entire data set as shown in
Fig.~\ref{fig:fullFit}. We emphasize that this fit is done with a
fixed amplitude for the Gaussian determined by a contamination level
of $^{\mathrm{nat}}$K of 13 ppb as quoted by the collaboration.  Only
five parameters were allowed to float in the fit: the centre of the
Gaussian; the resolution; the flat component; and the slope and
intercept of linear trend. This surprising result shows how volatile
the DM interpretation is to small changes in the background model and
it again emphasizes the need for a more thorough discussion of
backgrounds by the DAMA collaboration. In generating
Fig.~\ref{fig:total-fit} above we attempted to remain as accommodating
as possible and used the Gaussian curve quoted in~\cite{Nozzoli} for
modeling the bump and not the fit shown in Fig.~\ref{fig:fullFit}.

One may also wonder how does the fit actually look like when signal is
included. While the above discussion should make it clear that the
inclusion of any signal with a small modulation fraction leads to a
poor fit, we have included two simple examples to illustrate this
point directly. We chose two benchmark points corresponding to two
best-fit points of the DAMA-reported modulation amplitude:
spin-independent elastic scattering\footnote{The results obtained do
  not depend very strongly on the precise details of the model, but are mostly affected by  the modulation fraction.} for low-mass DM ($m_{\rm DM} = 12~\GeV$
and $\sigma_{\rm SI} = 2\times 10^{-40}$) as well as high-mass DM
($m_{\rm DM} = 60~\GeV$ and $\sigma_{\rm SI} = 8\times 10^{-42}$) for
halo parameters $\bar{v} = 220$ km/s and $v_{\rm esc} = 550$ km/s. The
unmodulated rate contributed by each of these signals is then included
in the fit of the total rate. The results are shown in
Fig.~\ref{fig:fullFit}. The high-mass point is in complete
disagreement with the data as can be expected since the modulation
fraction in this case is small ($\sim 5\%$) and hence the contribution to the unmodulated rate is much too large. The low-mass benchmark is
in slightly better agreement with the data since it enjoys a somewhat
larger modulation fraction ($\sim 8\%$). However, it results in a very low flat component of $\sim 0.55$ cpd/kg/keV, which is in disagreement with the higher energy part of the unmodulated spectrum as discussed in the main text. This is a general feature of any signal for which the unmodulated component contributes significantly in the 2-4 keV region. On the other hand, a signal with a large modulation fraction ($\gtrsim 20\%$) would contribute very little to the total rate and can result in a satisfactory fit in accord with our original claim.

Finally, we stress that the purpose of the background analysis we
present is not to claim a full understanding of the backgrounds. Our
goal, as it was in our paper, is to call attention to this important
background and encourage the collaboration to present a full account
of its detailed understanding. That being said, the remarkable
agreement of our simple background model with the data must give pause
to anyone who wishes to interpret the DAMA results in terms of a
signal of dark matter.

\bibliographystyle{unsrt}
\bibliography{K40bib}
\end{document}